\documentclass{lhep}
\usepackage{graphicx}

\journal{LHEP}
\vol{xx}
\jyear{2022}
\pages{xxx} 

\received{30 October 2022}
\published{xx xxxxxx xxxx}

\def\be{\begin{equation}}
\def\ee{\end{equation}}
\def\bea{\begin{eqnarray}}
\def\eea{\end{eqnarray}}

\begin{document}

\title{High Energy Proton-Proton Interactions and Baryonium Dark Matter}

\author {O.I. Piskounova,\auno{1}}
\address {{$^1$} P.N. Lebedev Physical Institute of RAS, 119991 Moscow, Russia}

\begin{abstract}
The positive hyperon production asymmetries that have been measured at LHC are real demonstrations of string junction role in the baryon charge transfer at baryon production in HE proton-proton interactions. In order to invent the neutral heavy particle with zero baryon charge as a candidate for Dark Matter, it is necessary to turn back to the times of dual topological unitarization of hadroproduction physics, where  the pomeron exchanges play leading role with the growing energies. The topological presentation of pomeron exchange at the HE proton-proton collision is a cylinder that is covered with a quark-gluon net. Taking into account that the junction of three gluons (SJ) has the positive baryon number, as well as the antijunction (antiSJ) is of negative baryon charge, the neutral self-connected baryonium configuration with zero baryon charge is a torus that is covered by a discrete number of hexagons with 3 string junction and 3 antijunction vertices each. The possibilities to observe such baryonium tori are discussed in this paper for three cases: collisions at p-p collider, the cosmic particle collisions with the atmosphere, and in the relativistic jets at the Active Galaxy Nuclei as well.  
\end{abstract}

\maketitle


\section{INTRODUCTION}

Historically, the term baryonium appears in 1970th, based on the experimental expectations \cite{rosner}. The definition was done in \cite{chew}: "it was quickly realized by J.Rosner that duality for baryons implies new particles family - combination of two quarks with two antiquarks. Assuming these new states to be long-sought Rosner exotics, they have tentatively dubbed $"baryonium"$." The main feature of baryonium was marked in \cite{rosenzweig}:" The most striking property of baryonium states is the existence of selection rules, which forbids baryonium decay into purely mesonic states." This actually means the absence of annihilation between baryon and antibaryon. Same time, the string interpretation of dual topological models was developed \cite{veneziano} that led to topological expansion of pomeron diagrams, which are responsible for HE hadron production. The studies on various colliders \cite{SJtransfer,arakelyan} have shown that the baryon charge is transferred by string junction (SJ), the configuration of three gluons, which is connecting three quarks in the baryon and, correspondingly, the same for antiSJ  in antibaryon. The purpose of this work is to build the objects from SJs and antiSJs, which are self-connected, have zero baryon charge, and are of valuable masses, in order to present them as baryonium DM candidates.  

\section{STRUCTURE AND MASSES OF BARYONIUM TOROIDAL PARTICLES}

\begin{figure}[h]
\begin{center}
\includegraphics[scale=0.42]{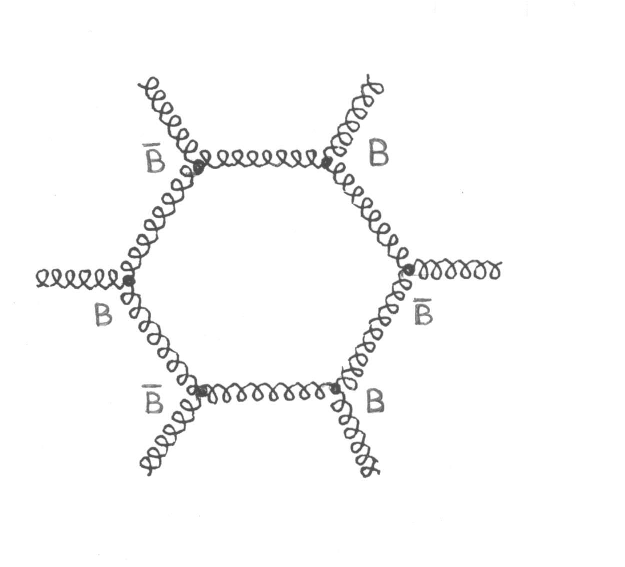}
\caption{ The hexagon diagram for the connection of 3 SJs and 3 antiSJs with zero baryon charge.  
\label{fig:hexagon} }
\end{center}
\end{figure}

Structure of baryonium DM particles was discussed in \cite{torus}. The idea was based on the hexagon diagram with three proton SJ and three antiproton ones, see Figure ~\ref{fig:hexagon}. In such a way, this construction is neutral with zero baryon charge. The hexagons can be collected into the net, and this hexagon net is able to be self-connected. It can be connected only on a torus surface. The number of hexagons is discrete, the number of gluon vertices is $N_{SJ+antSJ}$ = $2N_{hexagon}$. In this hypothesis, the baryonium means the nucleon-antinucleon construction with one parameter, it's mass that depends on the number of hidden baryon charge: $N_{SJ+antiSJ}$. The mass of baryonium DM particle  is discrete because of a discrete number of hexagons, which can be closed on the torus.

\begin{figure}[h]
\begin{center}
\includegraphics[scale=0.26]{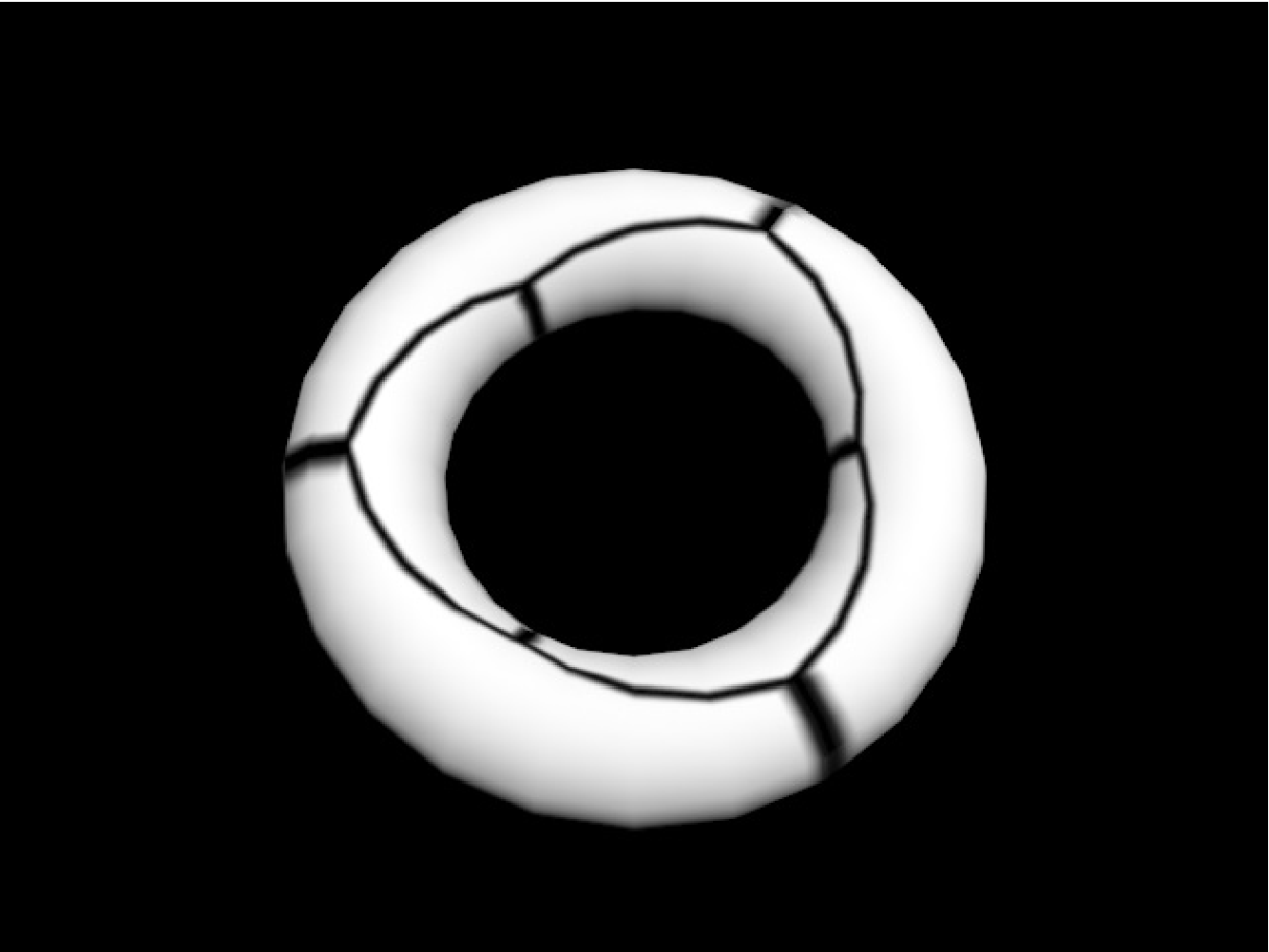}
\caption{ The 6 hexagon net and torus that is covered with the net of 12 connected  SJ-antiSJ vertices .  
\label{fig:torus} }
\end{center}
\end{figure}

The self-connected SJ+antiSJ construction, see Figure ~\ref{fig:torus}, certainly has discrete  progression of masses. 

The invisible neutral states of hadron matter have been studied in \cite{ptvsenergy, ptvsmass}, where the masses of possible pre-hadronic neutral QCD states go with the progression: $M_n = 0.251*e^{n-1}$ GeV. The lowest mass can be interpreted as a construction of just one SJ with one antiSJ. The first state with a mass that is equal to 0.25 GeV should be the proton-antiproton SJ-antiSJ state, which has reduced its mass almost 8 times. Such neutral baryon-antibaryon construction can be incorporated into the nucleon. 
     
\section{{TOROIDAL BARIONIUM DM}}

Let us consider where the baryonium Dark Matter (BDM) can be met. First of all, it should participate in multipomeron exchanges in proton-proton collisions at LHC and certainly play an important role in the relativistic jets from Super Massive Black Holes (SMBHs).
There are also a lot of puzzles and mysteries in the physics of cosmic rays that could be explained by BDM collision with the atmosphere.

\subsection*{BDM at LHC}

\begin{figure}[h]
\begin{center}
\includegraphics[scale=0.34]{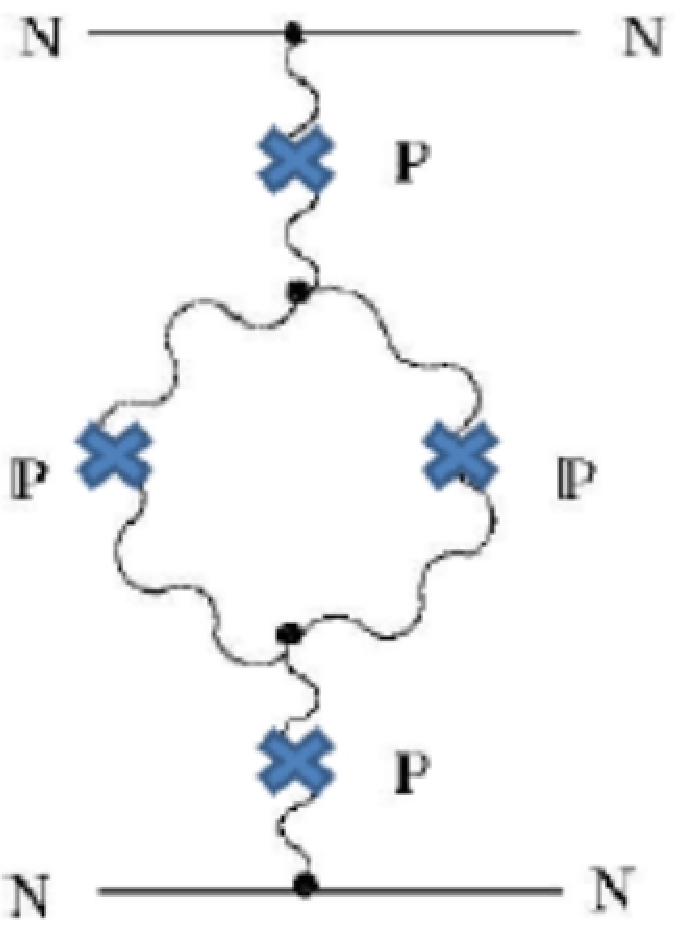}
\caption{ The pomeron loop in pomeron-with-handle order of topological expansion: the crosses mean the possible cuts. 
\label{fig:pomeronloop} }
\end{center}
\end{figure}

Double diffraction dissociation (DD) is a next-to-leading order contribution in the topological expansion after the pomeron exchange and should be presented as one pomeron diagram with a pomeron loop in the center, see Figure ~\ref{fig:pomeronloop}. Actually, the DD configuration is similar to a cylinder with a handle that takes $1/{{N_f}^4}$ part from the single pomeron exchange cross section that is 1.2 percent of $\sigma_{inel}$. Looking at the pomeron loop in this diagram, we are realizing that it is SJ-antiSJ torus, or BDM, in 3D topology. 

If the central pomeron loop (PL) is not cut, we have the DD spectra of produced hadrons: two intervals at the ends of a rapidity range, which are populated with particles of multiple production, and the valuable gap in the center of rapidity. Otherwise, if both sides of PL are cut, we have the same gap in the center that is populated with produced hadrons of doubled multiplicity. Both signatures together would give 2.4 percents of probability of detection the toroidal hadron matter in the processes of particle production at HE proton-proton colliders.

\subsection*{BDM at Cosmic Ray collisions with atmosphere}

The analysis of a hadroproduction event of 1975  in the stratosphere has been carried out recently \cite{stratosphere}. The rapidity and transverse mass distributions have been compared with the spectra that are measured in nucleus-nucleus collision at LHC. The conclusions have been as follow: 1) The value of central multiplicity corresponds to carbon nucleus (12 nucleons) collision with the typical carbon-nitrogen-oxygen (CNO) nucleus of the atmosphere.
The maximal rapidity corresponds to the energy per nucleon of the order of 500 GeV in the center-of-mass system. 2) Two leading particles have transverse mass near 1 GeV and are interpreted as protons that contribute into the visible peak at the end of a rapidity spectrum. Two protons in this peak hardly correspond to carbon collision. 3) One track with $M_t$ = $\sqrt{M_0^2+P_t^2}$ = 16 GeV shows the outstanding value of transverse mass $M_t$ that is far from the mass range of all other well-known hadrons. This makes us conclude that a new state of hadron matter can exist with a mass of approximately 14 GeV. These heavy neutral hadronic states \cite{torus} are good candidates for Dark Matter. These states have the Regge type of component distribution, as it is seen from the rapidity histogram, which leads to a small amount of protons in the fragmentation region. Moreover, this sort of QCD construction can split into the pair of similar ones with lower mass, so two tracks of hadrons with the suprisingly heavy masses could be seen.

\subsection*{BDM in relativistic jets}

Manifestations of toroidally organized matter are typical for space observations: remnants of Supernovae, galactic dust, etc. 
There are few observations of possible BDM near the Super massive Black Holes (SMBHs). As an example, here are two of them: a) dense matter torus arround of SMBH, see \cite{chandra},  and b) inside the relativistic jets radiated from SMBH \cite{radiojets}. 
The last gives a key to understanding how the relativistic jets are created.
The toroidal BDM, which was injected from SMBH with a jet, is invisible in the beginning because a baryonium torus has to  inflate moving out of compressing gravitation area. And then the deconstruction of such matter leads to a visible particle shower. Initially, the BDM torus has the giant potential energy of self-connection. This energy may go for the acceleration of produced particles \cite{protonacceleration}. Such is the new concept of particle production with the relativistic jets from Active Nuclei Galaxies.

\section{{CONCLUSIONS}}

Baryons have always been the most complicated and challenging objects in the theory of hadron interactions \cite{rosner}. The recent result of baryon hadroproduction studies \cite{SJtransfer,arakelyan} is the convincing implication that string junction of proton transfers baryon charge to the central kinematical region at HE proton collisions. It is suggested that SJs can build, at some extremal conditions, Dark Matter particles \cite{torus}. These particles have been named baryonium Dark Matter (BDM), because they have zero baryon charge. Their structure includes SJs and antiSJs symmetrically. BDM may be stable in strong gravitation environments. If very massive, it can be stable in a vacuum and be spread in the space. It was observed in \cite{ptvsmass} that the hadron mass generations may be the result of a decay of neutral hadron states with the geometric progression of masses. The infinite mass scale of BDM particles gives us the possibility to fit the data on DM in space with the parameters of a BDM mass population. BDM particles of great mass are to be injected with relativistic jets from SMBHs. It can be assumed that the potential energy of self-connected baryonium toroids, after their disintegration, gives a valuable acceleration to protons in jets \cite{protonacceleration}. The surviving  low-mass BDM particles will split through collisions with the nuclei of the atmosphere  into the pair of lower-mass BDMs, see \cite{stratosphere}. At the collider experiments, baryon-antibaryon toroid is just a third-order term in the topological expansion of pomeron exchange amplitudes, which takes place in 1.2 percent of inclusive production events. It will take time to answer new multiple questions that are arising from the baryonium DM concept. Finally, this approach suggests that baryon matter and dark matter can overflow one into another.

\bibliographystyle{unsrt}

\begin{thebibliography}{99}
\bibitem{rosner} J.L. Rosner, Phys. Rev. Lett. {\bf 21} (1968), 950.
\bibitem{chew} G. Chew and C. Rosenzweig, Phys. Rep. {\bf 41} C (1978), 263.
\bibitem{rosenzweig} C. Rosenzweig, Phys. Lett. {\bf 71}B (1977), 203.
\bibitem{veneziano} K. Giafaloni , G. Machesini  and G. Veneziano , Nucl. Phys. B {\bf 96} (1975), 472.
\bibitem{SJtransfer} O. I. Piskounova,  Phys. Atom. Nucl. {\bf 70} (2007), 1107, [arXiv:hep-ph/0604157].
\bibitem{arakelyan} G.H. Arakelyan {\it et al.} [arXiv:0709.3174].
\bibitem{torus} O. I. Piskounova, [arXiv:1812.02691].
\bibitem{ptvsenergy} O.I. Piskounova, Int. Jou. of Mod. Phys. A {\bf 35} (2020), 2050067, [arXiv:1706.07648].
\bibitem{ptvsmass}O.I. Piskounova, [arXiv:1908.10759].
\bibitem {stratosphere} O.I. Piskounova, PoS(ICHEP2022), {\bf 910}, [arXiv:1907.00176].
\bibitem{protonacceleration} O.I. Piskounova, PoS(ICHEP2022), {\bf 911}, [arXiv:2110.13618 and 2302.08546].
\bibitem{chandra} B. Luo {\it et al.}, ApJ {\bf 805} (2015), 122, [arXiv:1503.02085]
\bibitem {radiojets} J. Rawes , M. Birkinshaw  and D. Worrall, Mont. Not. of RAS, {\bf 480} ( 2018), 3644, [arXiv:1808.01967]. 

\end{thebibliography}

\end{document}